\documentclass[twocolumn,showpacs,preprintnumbers,amsmath,amssymb]{revtex4}

\usepackage{graphicx}

\begin{document}

\title
{ The electron effective mass in the 
strongly correlated  2D-uniform electron fluid from
finite-temperature calculations.
}

\author
{
 M.W.C. Dharma-wardana}
\affiliation{Institute of Microstructural Sciences,
National Research Council of Canada, Ottawa, Canada. K1A 0R6\\
}
\email[Email address:\ ]{chandre.dharma-wardana@nrc-cnrc.gc.ca}

\date{\today}
\begin{abstract}
The  very-low temperature thermal effective mass $m^*$ of  paramagnetic and
ferromagnetic  electrons in a uniform electron fluid in two dimensions is studied.
Analytical and numerical evaluations are used to meaningfully define an $m^*$ even in
the the Hartree-Fock approximation. The Hartree-Fock $m^*$ decreases linearly with the
electron-disk radius $r_s$. Correlation effects lead to strong cancellations between
exchange and correlation. Thus the effective mass is enhanced with increasing $r_s$
for  the unpolarized fluid, while $m^*$ decreases with the $r_s$ of
 the polarized fluid.
The  effective mass is calculated from the coefficient of the
quadratic temperature dependence of exchange-correlation free energy $F_{xc}$. This is
calculated in a physically transparent manner using a new formula for the effective
mass. This uses the $T=0$ pair-distribution functions of Gori-Giorgi et al., and the temperature derivative of a quantum analogue of the potential of mean-force well known in the statistical
mechanics of classical fluids. The results are compared with recent quantum
Monte-Carlo simulations at $T=0$, as well as with other available experimental and
theoretical  data for the effective mass.
\end{abstract}
%
\maketitle

\section{Introduction.}
At very low temperatures, essentially at $T=0$, the electrons in a 2D uniform electron
fluid (UEF) occupy momentum states  up to the Fermi energy $E_F$.  If the electrons
were non-interacting they have only kinetic energy, the potential energy being a
constant. The energy bands of the electrons are of the form $k^2/2m$, where
the mass $m$ of the non-interacting electrons is unity, even in semi-conductor
nanostructures or metals where the material parameters can be subsumed by a suitable
re-definition of the effective atomic units~\cite{afs}. Such 2D electron layers have
been experimentally realized in semiconductor interfaces, enabling the experimental
probing of such systems which are of great fundamental and technical importance.

If the the number of electrons in a unit area (measured in atomic units) is $n$, the
radius of the disk in real space per electron, {\it viz}. $r_s$, is $1/\surd(\pi
n)$. The ratio of the Coulomb interaction to the kinetic energy is a measure of the
`strength'  of the interactions in the system. In the limit $T\to 0$  it is
found to be equal to $r_s$.
Thus $r_s$ plays the role of the 'expansion parameter' or coupling constant in the
perturbation theories of the electron fluid. Perturbation theories work reliably when
the coupling $r_s$ is less than unity. The range of validity can sometimes be increased
by the use of a `screened interactions', at least for some properties.

One of the tenets of Fermi liquid theory, originally developed to treat the behaviour
of  electrons in metals (where usually $2 < r_s<5$ ),  is that the  low-energy
excitations in these interacting systems, known as quasiparticles, are very similar
to  non-interacting electrons. However, the mass $m$ is replaced by an effective mass
$m^*$ which allows for the modification of the energy bands due to many-body
interactions. Other parameters which describe the electrons, e.g., the Land\'{e} $g$
factor, are also modified and these are the `Fermi-liquid parameters' of Landau
theory. Unfortunately, standard many-body theory which uses perturbation methods
cannot make a reliable evaluation  of $m^*$ for systems where $r_s$ is greater than
unity. Physically motivated approximations are needed to truncated the perturbation
chain, and these invariably result in the failure to satisfy the sum rules,
self-consistency conditions etc. If the results were insensitive to the various
possible choices for screening, vertex corrections, selection of graphs, etc., then
this would not matter. Unfortunately, the evaluation of $m^*$, usually carried out
from a physically motivated approximation to the self energy, turns out to be very
sensitive to the model used. Thus. for example,  perturbation theories may predict the
$m^*$ of electrons at $r_s$=5 to be a fraction of the bare mass, or  several times the
bare mass, depending on the model used. These methods even fail to predict
positive-definite pair-distribution functions (PDFs), i.e., $g(r)$ 
for useful values of $r_s$.

However, significantly more reliable results are available from quantum Monte Carlo
simulations (QMC), especially for PDFs and correlation energies. Unfortunately,  the
$m^*$  evaluations involve the probing of excited states, and  only a few QMC
calculations are available for the effective mass of the 2D UEF. Experimentally too,
 the measurements have
been very challenging. Nevertheless, currently available results, both experimental
and simulational, show that the $m^*$ of the 2D-paramagnetic electron liquid is
enhanced above unity as $r_s$ increases, while the  $m^*$ of the fully spin-polarized 
system remains below unity even as $r_s$ is increased.

 An alternative approach to the study of the effective mass is to look at
quasiparticle excitations in a fluid at a temperature $T$ close to $T=0$. The thermal
excitations occur in a strip of energy of width $T$ near $E_F$.  Th excitations in the
system are associated with an increase in the Helmholtz free energy $F_0$
(non-interacting case), or $F$ (interacting case), of the system. The quasiparticle
mass enters directly into the specific heat of the system. Hence the ratio of the
specific heats of the interacting system and the non-interacting system provides a
direct and unambiguous measure of the effective mass $m^*$ of the excitations. Under
certain conditions, these can be identified with the Landau quasiparticles of the
Fermi liquid in the limit $T\to 0$  as discussed by Luttinger and others
\cite{Luttinger60}. Thus a calculation of the interacting free energy $F$ as a
function of $r_s$ at finite-$T$ would provide an estimate of the effective mass $m^*$,
if the finite-$T$ free-energy of the system could be calculated.

The calculation of the finite-$T$ interacting free energy or the self-energy at the
Hartree-Fock (HF) level is well controlled. However, the self-energy is divergent at
the Fermi energy and it is not possible to define an effective mass. On the other
hand, the logarithmic divergencies in the HF-free energy can be separated out and we are
able to present a thermal effective mass at the HF level, viz., $m_x^*$, arising
entirely from exchange processes.  Here we present previously unpublished results for
the coefficients of the $T^2$ term of the polarized system. The HF $m_x^*$ decreases
with $r_s$  and this approximation breaks down for  $r_s > 3$  when $m_x$ becomes
negative. Thus the inclusion of correlation corrections is imperative to obtain a
physically meaningful result.   

Instead of using diagrammatic methods, in our previous work we used a calculation of
$F_{xc}$ at finite-$T$ {\it via} a coupling-constant integration of the PDFs of the
interacting system. The finite-$T$ $g(r)$ needed for the calculations were obtained
using the classical-map hyper-netted-chain (CHNC) method where the 2D electron system
at the temperature $T=0$ is replaced by a classical Coulomb fluid at the temperature
$T_q$.  Given that the interacting chemical potential $\mu$  becomes negative beyond
$r_s\sim2$, (e.g., at $r_s=5$, $\mu/E_F\sim -5$), the distribution functions even at
$T$ just slightly above $T=0$ are classical Boltzmann distributions with little or no
occupation at $k_F$. Hence the study of a classical model which correctly
incorporates quantum features via effective potentials is a very reasonable
proposition. However, although the CHNC can be unambiguously implemented at $T=0$ and
at sufficiently elevated temperatures, we are beset with a number of difficulties in
dealing with the `warm-dense' region close to $T=0$. The two main difficulties
are (a) the elimination of logarithmic-divergent terms which need to exactly cancel
with the exchange and correlation contributions, (b) the finite-$T$ 
modeling of the bridge
function of the classical-fluid which controls the cluster diagrams beyond the
hyper-netted-chain sum of diagrams. In this study we present an alternative approach
which partially circumvents these difficulties, and provides a more transparent
analysis, leading to a new formulation of the effective mass calculation.

The $g(r_s,r)$, i.e.,  PDFs of the 2D-electron system (at $T=0$) have been accurately
parametrized by Giri-Giorgi et al., \cite{ggpair} and may be considered known. The
$F_{xc}$ at $T=0$ can be written as a coupling-constant integration over $g(\lambda
r_s, r)$ where $\lambda$ is the coupling constant. Further more, we consider
-$log\{g(r)\}$ as the potential of mean force, viz., $\beta V_{mf}(r)$  of the
equivalent classical interacting Coulomb fluid. Then we determine the finite-$T$ form
of  $\beta V_{mf}(r)$ to second order in the temperature and use this to directly
evaluate the second-order temperature correction to the exchange-correlation free
energy. In the following we show that the method leads to a transparent, if
approximate, calculation of the effective mass $m^*$ at arbitrary polarizations, and
in good agreement with the available results in the field.
\section{Theory}
\label{theory-sec}
The thermal effective mass $m^*$ can be expressed as a ratio of the
heat capacities of the 
interacting and non-interacting systems as:  
\begin{equation}
\label{mstareq}
m^*=C_v/C_v^0=\frac{\left[\partial^2 F(T)/\partial T^2\right]}
{\left[\partial^2 F_0(T)/\partial T^2\right]}
\end{equation}
Here $C_v$ is the specific heat at constant volume. The interacting free energy $F$ is
the sum $F_0+F_x+F_c$, where $F_x$ and $F_c$ are the exchange and correlation
contributions, with $F_{xc}=F_x+F_c$.
 Hence the problem of determining $m^*$ reduces to a calculation of
exchange-correlation effects at finite temperatures, near $T=0$. However, such
calculations are in many ways even more demanding than those at zero temperature, as
perturbation methods have to now deal with a whole host of new diagrams, their
singularities and cancellations. These difficulties were first addressed in the papers
by Luttinger, Ward, and Kohn \cite{Luttinger60}. Similarly, QMC methods are also
equally difficult, especially for $T$ very close to zero. 

The Hartree-Fock self-energy becomes logarithmically divergent near $k_F$ and it is
not possible to define an $m^*$ via the self-energy at $T=0$. In the following we
first examine the exchange-only, i.e., Hartree-Fock,  approximation to $F$ at finite
temperature, and calculate a {\it regularized} effective mass $m_x^*=m_{HF}=1+\Delta
m_x$ for paramagnetic and ferromagnetic 2D electrons, i.e., for spin polarizations
$\zeta=0$ and 1. It is found that $m_x$, containing the corrections from $F_x$, is a
decreasing linear function of $r_s$ for both polarizations. 

\subsection{Non-interacting and Hartree-Fock Helmholtz free energies}
\label{hf-sub}
The non-interacting free energy $F_0=E_0-TS$, where $S$ is the entropy, can be
evaluated and expressed in terms of the dilogarithm function. Writing $\eta=\mu/T$,
where $\mu$ is the chemical potential, $F_0$ and $E_0$ for the 2D system are given as
follows,  per unit atomic volume, and in Hartree atomic units:
\begin{eqnarray}
\label{e0f0-eq}
E_0&=&\frac{T^2}{2\pi}\left[\frac{\pi^2}{6}+\frac{\eta^2}{2}+
\mbox{dilog}(\frac{e^{\eta}}{1+e^\eta})\right]\\ 
F_0&=&\mu n-\frac{T^2}{2\pi}\left[\frac{\pi^2}{6}+\frac{\eta^2}{2}+
\mbox{dilog}(1+e^{-\eta})\right]
\end{eqnarray}
Alternative expressions are possible, as given in ref~\cite{pd2d, prl2}.
The low-temperature expansion of $F_0(t)$ can be given as:
\begin{eqnarray}
\label{f0-eqn}
F_0(t,\zeta=0)&=&E_0(0,0)(1-3.2859t^2)\;\;\;\zeta=0 \\
F_0(t,\zeta=1)&=&E_0(0,1)(1-0.82246t^2)\;\;\;\zeta=1 \\
E_0(0,\zeta)/n&=&0.5(1+\zeta^2)/r_s^2
\end{eqnarray}
The quadratic coefficient in the expansion of $F_0(t,\zeta)$ would be
denoted by $a_2(\zeta)$ where needed. Thus $a_2(\zeta=1)=-0.82246E_0$.
Note that here we have used the same paramagnetic $E_F$ in defining $t=T/E_F$
for all 2D systems mentioned in the above equations.

The first-order (i.e., unscreened) exchange free energy $F_x$ consists
of  $F_x^i$, where $i$ runs over the
two spin species. At $T=0$ these reduce to the exchange
energies:
\begin{equation}
E_i^x/n=-\frac{8}{3\surd{\pi}}n_i^{1/2}
\end{equation}
Here $n_1=n(1+\zeta)/2$, and $n_2=n(1-\zeta)/2$.
Then the exchange energy per particle at $T=0$, i.e., the internal energy
 contribution $E_x/n$
at $T=0$ becomes
\begin{equation}
\label{ex_t=0}
E_x/n=(E_1^x+E_2^x)/n=-\frac{8}{3\pi r_s}[c_1^{3/2}+c_2^{3/2}]
\end{equation}
Here $c_1$ and $c_2$ are the fractional compositions
$(1\pm\zeta)/2$ of the two spin species.

We also define the species-dependent
reduced chemical potentials  $\mu^0_i/T$ by $\eta_i$,
 reduced
temperatures $t_1=t/(1+\zeta)$ and $t_2=t/(1-\zeta)$, based on the
two Fermi energies $E_{F1}$ and $E_{F2}$ which are $E_F(1\pm\zeta)$.
Then we have:
\begin{equation}
\label{exF}
F_i^x/E_i^x=\frac{3}{16}t_i^{3/2}\int_{-\infty}^{\eta_i}
\frac{I^2_{-1/2}(u)du}{(\eta_i-u)^{1/2}}
\end{equation}
The $I_{-1/2}$ is the Fermi integral defined as usual:
\begin{equation}
I_\nu(z)=\int_0^\infty\frac{dx x^\nu}{1+e^{x-z}}
\end{equation}
The $\eta_i$ are given by
\begin{equation}
\eta_i =\log(e^{1/t_i}-1)
\end{equation}
In the paramagnetic case
 Eq.~\ref{exF} reduces to the result given by
Isihara et al.~\cite{Isihara80} (see their Eqs.~3.4-3.6; they use
a slightly different definition of the Fermi
integral).  
 
The total exchange free energy per unit atomic volume
is $F_x=\Sigma F_i^x$.
The accurate numerical evaluation of Eq.~\ref{exF} requires the removal of
the square-root singularity by adding and subtracting, e.g.,
$I^2(-|\eta|)/(v-|\eta|)^{1/2}$ for the case where $\eta$ is negative, and
$v=u$, and so on. 

 A real-space formulation of $F_x$ = $F_1^x+F_2^x$
using the zeroth-order PDFs fits  naturally with
the approach of our study using pair-distribution functions of the
electron fluid as the main ingredient. Thus
\begin{equation}
\label{fxbyg0-eq}
F_x/n=n\int \frac{2\pi r dr}{r}\sum_{i<j}h^0_{ij}(r)
\end{equation}
Here  $h^0_{ij}(r)=g^0_{ij}(r)-1$.
In the non-interacting system at temperature $T$,
 the antiparallel $h^0_{12}$, viz.,
 $g_{12}^0(r,T)-1$,
 is zero while
$$h_{11}^0({\bf r}) 
=-\frac{1}{n_i^2}\Sigma_{{\bf k}_1,{\bf k}_2}n(k_1)n(k_2)
e^{i({\bf k}_1-{\bf k}_2){\bf \cdot}{\bf r}} \,\,
=\, -[f(r)]^2$$
Here {\bf k}, {\bf r} are 2-D vectors and $n(k)$ is
 the Fermi occupation number at
the temperature $T$. At $T=0$ $f(r)=2J_1(k_ir)/k_r$ where
$J_1(x)$ is a Bessel function. As a numerical check, we have evaluated the
exchange free energy by {\it both} methods, i.e., via $k$-space
and $r$-space calculations.

The following  small-$T$ expansions are useful for our purposes:
\begin{eqnarray}
\label{fxz0eq}
F_x(r_s,t,\zeta=0)&=&E_x(r_s,\zeta=0)[1+(\pi^2/16)t^2\log(t)\nonumber\\
 & &-0.56736t^2+\cdots]
\end{eqnarray}
This result, i.e., $F_x$ for the unpolarized system, has been given by
 Isihara et al.,~\cite{Isihara80} and re-confirmed by Mahan et al \cite{mahan}. The
corresponding internal energy, $E_x$ can be obtained from
the relation $E=d\{\beta F\}/d\beta$ where $\beta=1/T$. We
refer to the coefficient of the $t^2log(t)$ terms as $A_L$, and that of the quadratic
term as $A_2$. These depend on the spin polarization $\zeta$. The results for the
fully polarised case do not seem to be previously  available in the literature.
 We have obtained the following expansion:
\begin{eqnarray}
\label{fxz1eq}
F_x(r_s,t,\zeta=1)&=&E_x(r_s,\zeta=1)[1+(\pi^2/64)t^2\log(t)\nonumber \\
 & &-0.28368t^2+\cdots
\end{eqnarray}

\subsection{The Hartree-Fock effective mass}
\label{hfmstar-sub}
\begin{figure}
\includegraphics*[width=8.0 cm, height=8.0 cm]{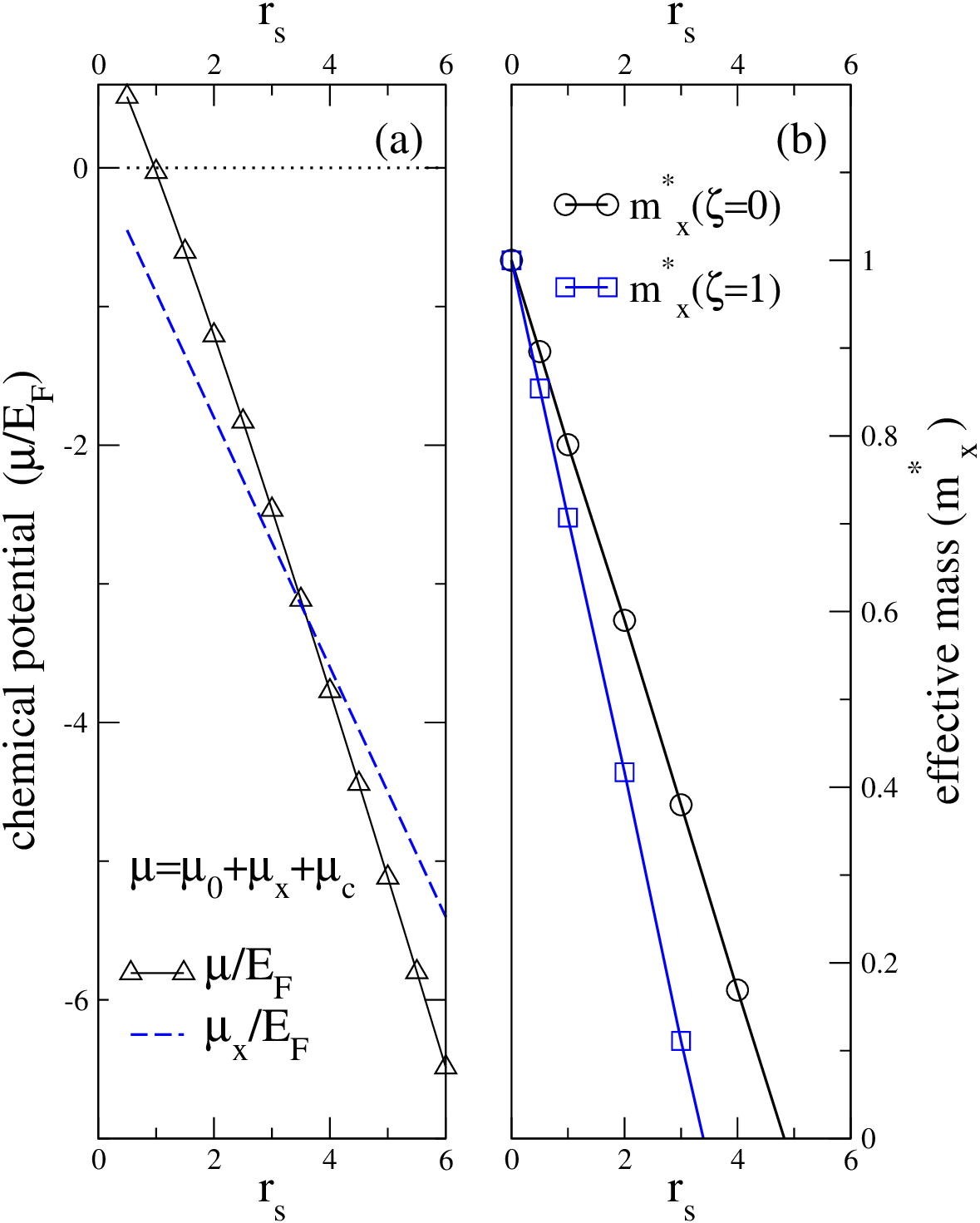}
\caption{
The panel (a) shows the modification of the chemical potential $\mu$ due to
interactions, driving it to the negative regime for $r_s>\sim 1$. Negative
$\mu$ is typical of classical fluids. (b) The Hartree-Fock effective mass,
 $m^*_x$, is shown
as a function of the density parameter
$r_s$, for the polarizations $\zeta=0$ and 1.
}
\label{mstarx-fig}
\end{figure}

The Hartree-Fock electron fluid is a theoretical construct which does not exist in
nature. However, it is a very useful conceptual model. The difficulties in this
conceptual model are due to the singular behaviour of the single-particle energy,
self-energy etc., close to the Fermi energy. Thus taking derivatives near the Fermi
energy becomes meaningless. However, as these divergences are logarithmic, they can be
integrated over and the free energy and related quantities can be evaluated.   They
are found to contain logarithmic terms which are explicitly exposed in
Eqs.~\ref{fxz0eq} and \ref{fxz1eq}. These logarithmic terms, of the form $A_L(\zeta)
t^2 log(t)$ are removed when higher order corrections are included in the theory,
while the quadratic term $A_2(\zeta) t^2$ contributes to the heat capacity.
 Hence it is clear
that we can define a {\it regularized} effective mass by dropping the $t^2log(t)$
terms and calculating an $m_x$ entirely from the coefficient $A_2$. Thus
\begin{equation}
\label{mx-eq}
m_x(\zeta)=1+A_2(\zeta)/a_2(\zeta)
\end{equation}
Here $a_2$ is the coefficient defined in Eq.~\ref{f0-eqn} in the expansion of $F_0$.
Results from such calculations are shown in Fig.~\ref{mstarx-fig}. These results show how
the exchange contributions widen the Hartree bandwidth with the decrease of the
effective mass. For $r_s >\,\sim 3$, the value of $m^*$ becomes negative.
 Clearly, the Hartree-Fock model is misleading except  at smaller $r_s$.
%
%
\section{Correlation corrections to the free energy and the effective mass.}
\label{mstar-corr-sec}
Microscopic theories of Landau Fermi liquids calculate  $m^*$  from the solutions of
the Dyson equation for the one-particle interacting Green's function of the system. If
the real part of the retarded self-energy is $\Sigma_1(\vec{k},\omega)$, the Landau
quasi-particle excitation energy $E_{QP}(\vec{k})$, measured with respect to the
chemical potential is used in calculating the effective mass $m^*$.
\begin{eqnarray}
E_{QP}(\vec{k})&=&\epsilon_k + \Sigma_1(\vec{k},\omega)|_{\omega=E_{QP}} \\
\epsilon_k &=& k^2/2-E_F\\
\frac{1}{m^*}&=&\frac{dE_{QP}(k)}{k_F\,dk}|_{k=k_F}
\end{eqnarray}
The quasiparticle energy is the eigenvalue of the Dyson equation and contains
the non-interacting energy $\epsilon(k)$ plus the real part of the the
selfenergy $\Sigma(k,\omega)$ evaluated self-consistently  at the quasiparticle
energy itself. We may rewrite the selfenergy as an exchange part, and a
correlation contribution:
\begin{equation}
\Sigma=\Sigma_x(k)+\Sigma_c(k,\omega).
\end{equation}
This should be evaluated self-consistently to satisfy sum rules, Ward
identities etc., to give a conserving approximation. In practice, these
self-consistency conditions have to be stringently satisfied if the logarithmic
singularities in $\Sigma_x$ are to be completely cancelled by corresponding
singular terms contained in $\Sigma_c$. Thus any approximate theory must
necessarily be on guard against spurious contamination from inadequate
cancellations. These could greatly enhance the calculated $m^*$, as is indeed
observed  in some RPA-type calculations. Further more, the interacting chemical
potential $\mu/E_F$ is very different from unity, and strongly negative. These
make the attempt to impose self-consistency a very daunting task.  In the
temperature theory, say Eq.~\ref{fxz0eq} of the previous sub-section,
these would manifest as spurious contributions to $m^*$ proportional to $1/t$ as $t\to
0$.

The exchange and correlation free energy $F_{xc}(r_s,t)$ at any temperature and
spin-polarization  can be evaluated from pair-distributions functions rather
than from self-energies or Green's functions.  The exchange free energy $F_x$
is just the contribution to a coupling constant integration (see below) over
the PDFs at zero coupling (see Eq.~\ref{fxbyg0-eq}).
 The $F_x$, evaluated from the non-interaction PDF,
$g^0(r)$ via Eq.~\ref{fxbyg0-eq} contains singular logarithmic terms.
These
singular terms at zero coupling are offset by the contributions from the rest
of the coupling-constant integration reaching out to full coupling. 
\begin{equation}
\label{cci-eq}
F_{xc}(rs,t)/n=\int_0^1d\lambda \frac{n}{2}\int \frac{2\pi r dr}{r}
\sum_{ij}c_ic_j(g_{ij}(r,\lambda)-1)
\end{equation}
Here $\lambda$ is the coupling constant. The $r_s$, and $t=T/E_F$ dependencies
in the $g_{ij}(\lambda, r)$ are not displaced for brevity. In our previous work
(e.g., ref.~\cite{quasi} and references there-in) the  needed $g_{ij}(r)$ are
calculated at any given temperature $T$ (including $T=0$) using the
CHNC method
for 2D systems ~\cite{prl2, pd2d}. In CHNC, the quantum fluid at $T=0$ is
replaced by a classical fluid at $T_q=1/\beta$.  At finite temperatures $T > 0$,
$\beta=1/\sqrt{T_q^2+T^2}$. Although this method gave reasonable results for
$m^*$ via Eq.~\ref{mstareq}, the question of the extension of model bridge
functions to finite-$T$, as well as the accuracy of the cancellation of
logarithmic terms which should cancel accurately remained troubling issues.
Numerical calculations very close to $T=0$ are also very susceptible to
difficulties due to the sharpness of the Fermi functions at very low $T$.

Accurate pair-distribution functions of the 2D-UEF at $T=0$ are now available
from QMC as well as from analytical representations developed by Gori-Giorgi et
al~\cite{ggpair}. We use these functions instead of those generated from
the CHNC, as this is equivalent to replacing the hard-disk bridge functions of the
CHNC description with the Coulomb bridge
function~\cite{br2d}.
 If the 2D-PDF, i.e., $g(r)$, obtained by these methods were
that of a classical fluid, then it would be of the form:
\begin{equation}
\label{meanforce-eq} 
g(r)\mapsto 
exp\left[-\beta V_{mf}(r)\right] 
\end{equation} 
$V_{mf}(r)$ is known as the potential of mean force, and is simply the
Kohn-Sham potential at $r$ in a classical fluid where one particle is
already at the origin.
The CHNC attempts to construct $\beta V_{mf}(r)$ directly from
the diffraction-corrected Coulomb potential $V_{dc}(r)$
and the Pauli-exclusion potential $P(r)\delta_{ss'}$
via the modified HNC equation. These potentials are discussed in
 greater detail in refs.~\cite{prl2, pd2d}.
\begin{equation}
\label{chnc-eq}
g(r)=exp\left[-\beta V_{dc}(r)-\beta P(r) + N(r)+B(r)\right] 
\end{equation} 
Here $N(r)$ is known as the nodal function, while $B(r)$ is the bridge
functions\cite{rosen} which bring in multi-particle clustering effects
which are outside the scope of the hyper-netted-chain diagrams. The
potentials $V_{dc}(r)$, and $P(r)$ are long-ranged,
while $N(r), B(r) $ contain
 many-body effects, and screening effects
which damp $V_{dc}(r), P(r)$,  so that
$g(r) \to 1$ for large $r$.

In the present paper we follow a different strategy. The potential of mean
force, $\beta V_{mf}(r,T)$ at finite-$T$ is developed as a Taylor expansion
around $T=0$, i.e., around $\beta=1/T_q$ and $F_{xc}$ is calculated using the
$T=0$ PDF. 
\begin{eqnarray}
\beta &=& 1/(T_q^2+T)^{1/2}\\
\beta V_{mf}(r,T)&=&\beta V_{mf}(r,0)+(T^2/2)\frac{\partial^2 \beta
V_{mf}(r,T)}{\partial T^2}|_{T=0}\nonumber\\
 & &  \\
F_{xc}(T)/n &=& F_{xc}(T=0)/n+ (T^2/2)\Delta \tilde{F}_{xc}\nonumber \\
\label{2ndorder-eq}
\Delta\tilde{F_{xc}}&= &(T^2/2)\int_0^1d\lambda
   \frac{n}{2}\int \frac{2\pi r dr}{r}\Gamma(r,T,\lambda)\\
\Gamma(r,T\lambda)&=&\sum_{ij}c_ic_jg_{ij}(r,\lambda)\frac{\partial^2 \beta
V_{mf}(r,T,\lambda)}{\partial T^2}|_{T=0}\nonumber
\end{eqnarray}
In calculating the second-order temperature derivative of the potential
of mean force, we use the CHNC form, Eq.~\ref{chnc-eq}, and construct an
approximate simplified from as a screened pair potential.
\begin{equation}
\label{scr-pair.eq}
\beta V_{mf}(r,T)\simeq\{\beta V_{dc}(r)+\beta P(r,T)\}e^{k_{sc}(r)} 
\end{equation}
The effect of the nodal term $N(r)$ and the bridge term of Eq.~\ref{chnc-eq} is
simplified and subsumed in the screening wavevector $k_{sc}$. At high densities, i.e., $r_s
\leq 1$,  electrostatic potentials
screen at the Thomas-Fermi value $2\pi n/E_F=2$, and drop to about
half this value at low densities. $\;$This is further discussed in
the section on results, given below. 
Also, the $T$-dependence or polarization dependence 
in the screening parameter is neglected in the present study.

\begin{figure}
\includegraphics*[width=8.0 cm, height=11.0 cm]{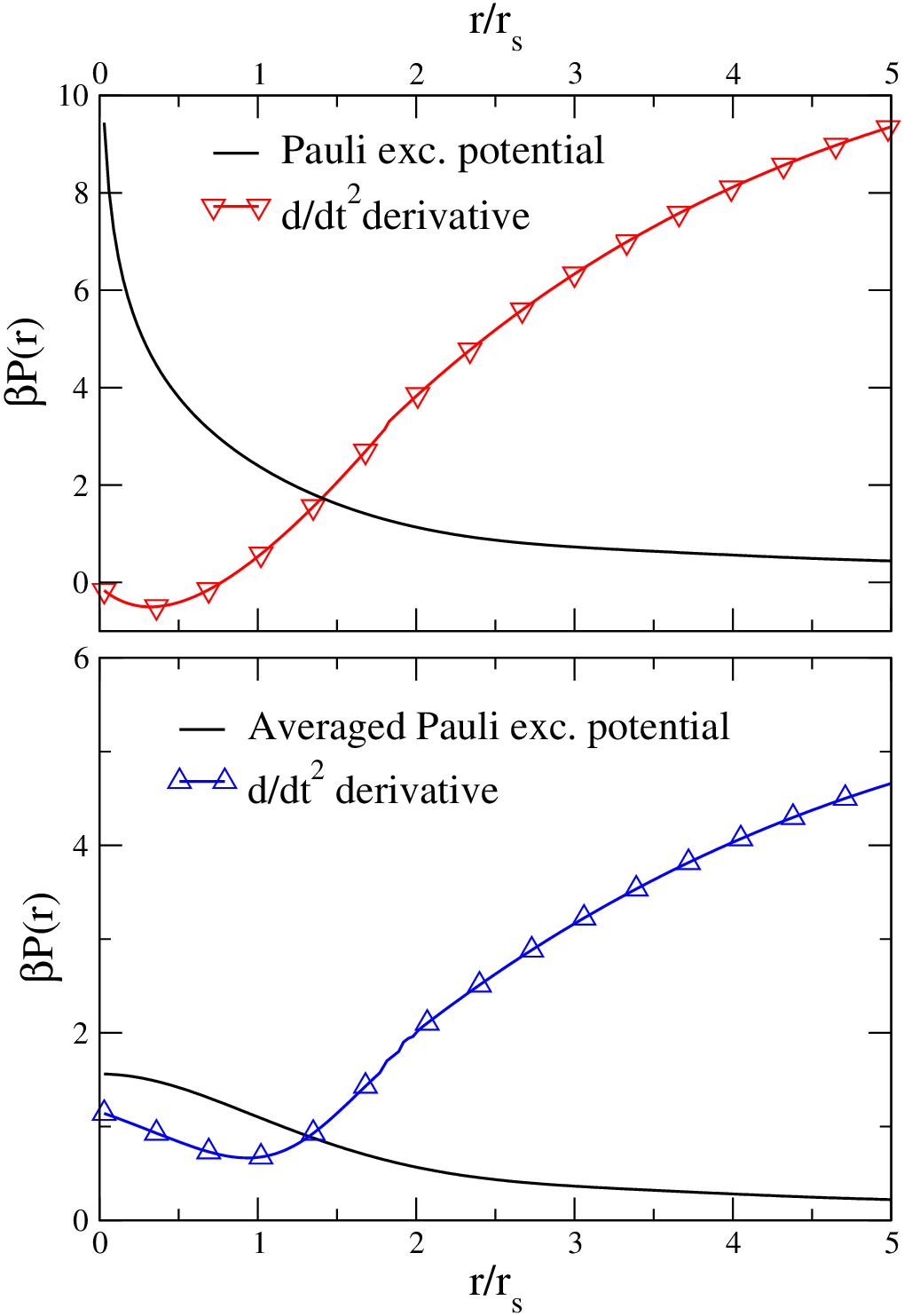}
\caption{The Pauli exclusion potential $\beta P(r)$ for
2D-electrons, and its second temperature derivative ($t=T/E_f$),
derived from $g^0_{11}(r)$. The bottom panel shows the results for the
paramagnetic, i.e., averaged $g^0(r)$. These potentials exactly reproduce
the Fermi hole in the pair-distribution functions.  
}
\label{pau-der.fig}
\end{figure}

The diffraction corrected  Coulomb potential and the Pauli exclusion
potential are of the form
\begin{equation}
V_{dc}= (1/r)\{1-e^{-k_{th}r}\}
\end{equation}
The de Broglie thermal wave vector $k_{th}=\{\pi T_q\}^{1/2}$ at $T=0$
appears in $V_{dc}(r)$. Its temperature dependence is neglected in our context 
as
it only modifies the very small-$r$ region, where $g(r)$ is negligible. Hence 
it plays no significant role in Eq.~\ref{2ndorder-eq}. 
The Pauli exclusion
potential $\beta P(r)$ is given by
\begin{equation}
\label{pauli-eq}
\beta P(r)\delta_{s,s'}=-log\{g^0(r)\}+(g^0(r)-1)+c^0(r) 
\end{equation}
Here $c^0(r)$ denotes the `direct-correlation function' 
 of Ornstein-Zernike theory. 
  It can be calculated from $g^0(r)$, i.e., the non-interacting
 pair-distribution function for parallel spins $s=s'$). An average potential
 may also be defined for use with the paramagnetic system treated
 as a single species, using $g^0_{av}=0.5(g^0_{11}+g^0_{12})$
 in Eq.~\ref{pauli-eq}.
  Hence $P(r)$ is known at any
 temperature and $r_s$. It is a universal function of $r/r_s$, and known
only as the product of $\beta$ and $P(r)$. However, we may use the
CHNC values of $T_q$ to define a $\beta$ at any $r_s$ and determine $P(r)$
by itself.  We have numerically calculated its
second-order temperature derivative (Fig.~\ref{pau-der.fig})
and denote it as $\beta P^{(2)}$ in the following.
These quantities are now assembled together to calculate the
second-order  finite-$T$ 
correction to $F_{xc}(\zeta)$. We have:
\begin{eqnarray}
\label{deltafx-eqn}
\Gamma(r,T\lambda)&=&\Gamma(mf)+\Gamma(P)\\
\Gamma(mf)&=&\sum_{ij}c_ic_jg_{ij}(r,\lambda)\frac{\beta V_{mf}(r,0)}{T_q^2}\\
\Gamma(P)&=&-\sum_{ij}c_ic_jg_{ij}(r,\lambda)\beta P^{(2)}(r,t)\delta_{ij}|_{t=0} \\
m^*(\zeta)&=&1+\Delta \tilde{F}_{xc}(\zeta)/a_2(\zeta)
\end{eqnarray}
The $a_2(\zeta)$ coefficient in the last equation is the coefficient of the
$T^2$ term in the interacting free energy, as in Eq.~\ref{f0-eqn}. The first
term in the r.h.s. of Eq.~\ref{deltafx-eqn} arises from the temperature
derivative of $\beta$ in $\beta V_{mf}(r, t)$, evaluated at $T=0$, while the
second term is from the second temperature derivative of the Pauli exclusion
potential. In evaluating Eq.~\ref{deltafx-eqn} we use the
pair-distribution functions of Giri-Giorgi et al. ~\cite{ggpair}, as the
approximate forms derived from CHNC  using a hard-disk bridge function
 are less accurate. The 2D 
quantum temperature $T_q$ and other CHNC procedures used here are those
given by Perrot and  Dharma-wardana in ref.~\cite{prl2}. 

\begin{figure}
\includegraphics*[width=8.0 cm, height=10.0 cm]{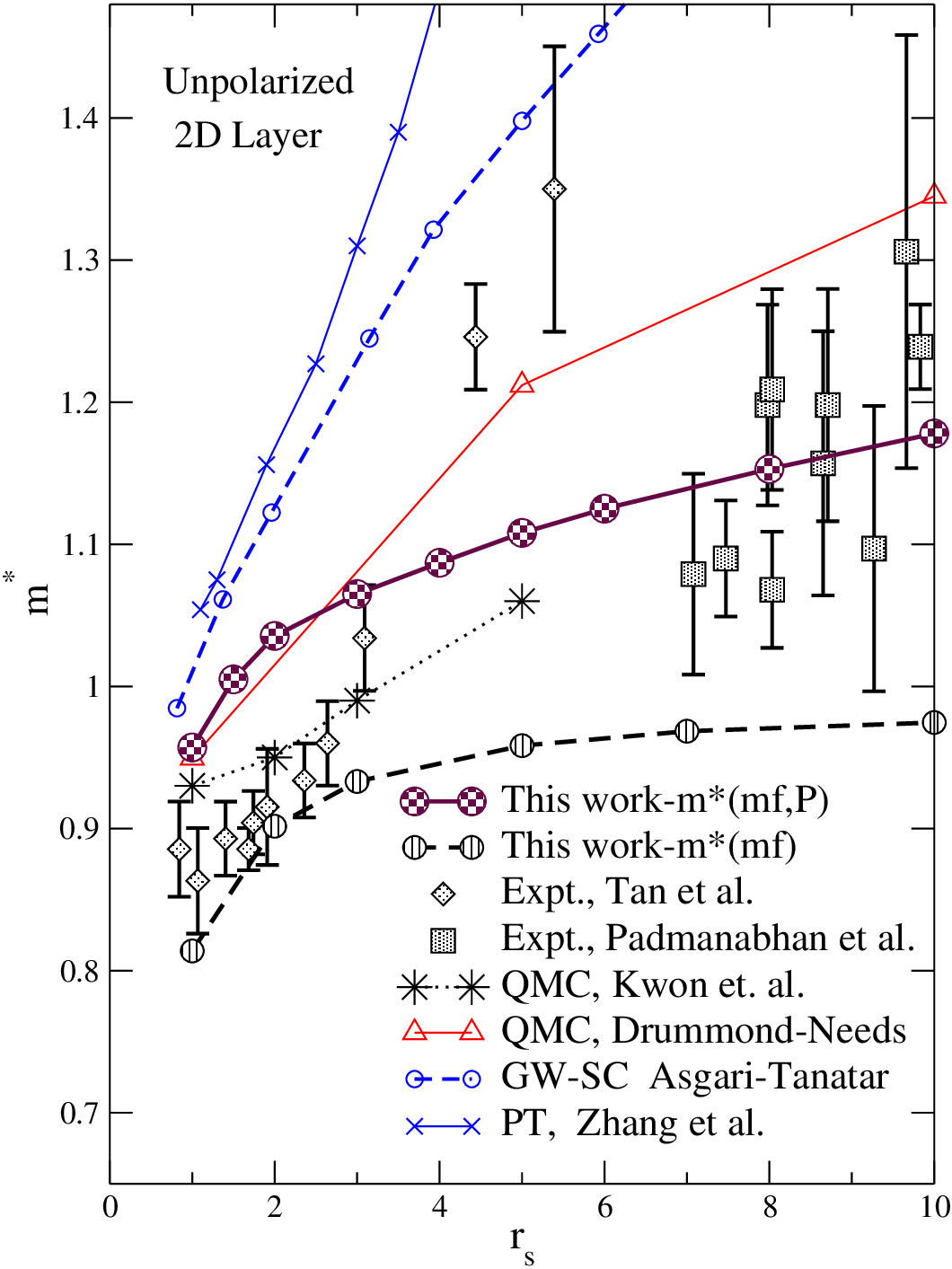}
\caption{
The effective mass $m^*$ as a function of the density parameter
 $r_s$, for unpolarized ($\zeta=0$) 2D electrons. See text
 for more details.  
}
\label{m2d-z0.fig}
\end{figure}

Eq.~\ref{deltafx-eqn} provides a new model for the calculation
 of the effective mass $m^*$ as a sum of
contributions from the temperature dependence of the potential of mean force,
$m_{xc}(mf)$, and
the temperature dependence of the Pauli-exclusion potential, $m_{xc}(P)$.
These contributions are shown in Figs.~\ref{m2d-z0.fig} and \ref{m2d-z1.fig}
and labeled according to the following equation.
\begin{eqnarray}
\label{mxc-def.eq}
m^*&=& 1+m_{xc}=1+m_{xc}(mf)+m_{xc}(P)\\
m^*&=&m(mf,P);\;\;m^*(mf)=1+m_{xc}(mf)\nonumber 
\end{eqnarray}
The contribution $m_{xc}(mf)$ may be thought of as resulting from temperature
dependent modifications
in exchange-correlation contributions as a function of $r_s$
due to screening, while $m(P)$ are contributions
from the screened exchange interactions. Only the second-order temperature derivatives have
been used as the first-order derivatives (which contain logarithmic terms of the form
$t^2log(t)$ are considered to cancel themselves out in an accurate theory.

%
%
%
\section{Results and discussion}
\label{results-sec}
\begin{figure}
\includegraphics*[width=8.0 cm, height=10.0 cm]{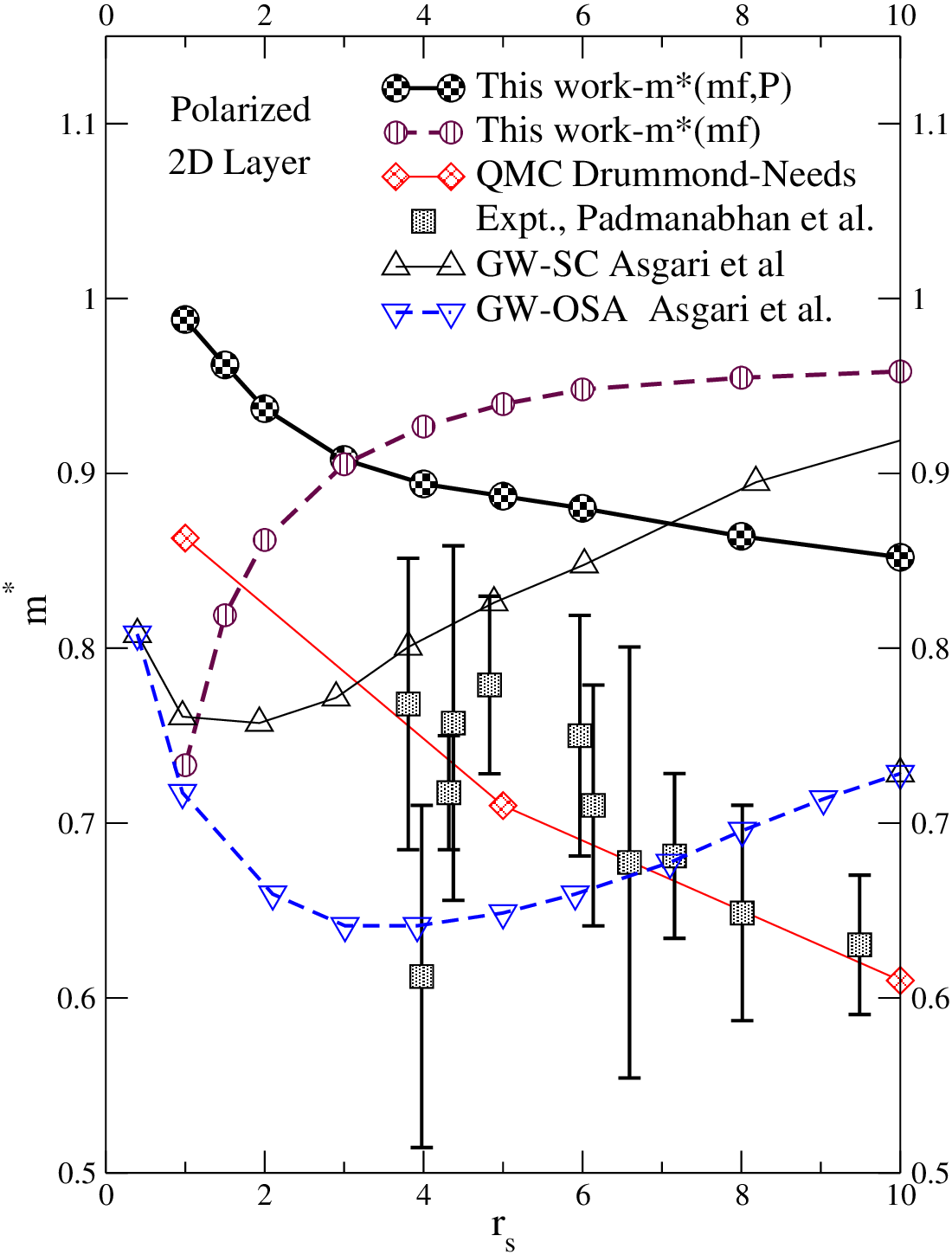}
\caption{
The effective mass $m^*$ as a function of the density parameter
 $r_s$, for polarized ($\zeta=1$) 2D electrons. See text
 for more details.  
}
\label{m2d-z1.fig}
\end{figure}

The values of $m^*$ calculated using different diagrammatic
perturbation expansions \cite{Asgari09,zhang} differ significantly
from each other, and from QMC 
results~\cite{kwon,Padman,DrumNeeds, Holzmann}.
However, the general trend seems to be that $m^*$ increases with $r_s$
for unpolarized ($\zeta=0$) electrons, while the $m^*$ for $\zeta=1$
decreases with $r_s$. We have summarized some representative results
as well as our results
in Fig.~\ref{m2d-z0.fig} for $\zeta=0$, and Fig.~\ref{m2d-z1.fig}
for fully polarized electrons.
Asgari et al. have given calculations where the Dyson equation
is solved ``self-consistently'' (GW-SC), as well as where an
``on-shell'' approximation (OSA) has been used~\cite{Asgari09}.
They have included approximate vertex corrections 
using the Kukkonenen-Overhauser (KO)
approach, but it is not clear if the fully interacting chemical
potentials (see Fig.~\ref{mstarx-fig}) have been used. 
 We  show the results
for $m^*(mf)$, as well as for the full $m*$ which includes the
contribution from the screened Pauli exclusion term. In the case
of $\zeta=0$ we have used the temperature derivative of
the averaged Pauli potentials (lower
panel, Fig.~\ref{pau-der.fig}, where as the
derivative of the full Pauli potential
was used with polarized electrons. Te temperature derivates were
calculated from numerical data in the range $T/E_F=0.05$ to $0.15$.
The essential weak point of this calculation is in the specifying of the
screening factor $k_{sc}$ in Eq.~\ref{scr-pair.eq}. A one-parameter screening
approach
is indeed very approximate. However, this
is guided by our calculation of $m^*(mf)$ shown in Fig.s~\ref{m2d-z0.fig}
and \ref{m2d-z1.fig}. While the high-density Thomas-Fermi $k_{sc}$ is
2 for electrostatic interactions, we have used 2.5 for $r_s$ below unity
for the screening of the exchange interactions brought in by the Pauli potential.
The screening parameter $k_{sc}$ drops to 2 at $r_s=4$, then to
$1.54$ at $r_s=10$, and unity near $r_s=30$. By $r_s=30$, $m^*$ becomes 1.34 for
 $\zeta$=0, and takes the value of 0.817 for $\zeta=1$. 
Drummond and Needs have recently reported
a QMC value of $m^*=1.34$ at $r_s=10$ for $\zeta=0$. Our model does not
exclude such higher values of $m^*$, but this would require a significantly
 weaker screening of the potentials, which accomodates less comfortably with
the behaviour of the potential of mean force (eq.~\ref{meanforce-eq} as a
function of $r_s$.

\section{Conclusion}
	We have presented results
for the effective mass  $m^*$ as a function of the density
 and polarization
 of a 2-D electron
fluid, using an entirely new approach based on the
known pair-distribution functions of the 2D fluid. 
Good agreement with the trends and magnitudes
observed in recent experimental and
theoretical approaches to the problem are recovered.
 The essential inputs to the present calculation
 are: (i) the pair distribution functions of the
2D-electron fluid as parametrized by Gori-Giorgi et al~\cite{ggpair},
(ii) the Pauli exclusion potential extracted from the non-interacting PDF,
(iii) elementary thermal physics, and (iv) some classical-map HNC
 concepts from the theory of classical statistical mechanics. The method
 provides a
physical understanding of the processes contributing to the enhancement or
diminution of the effective mass.


\begin{thebibliography}{99}

\bibitem{afs}
T. Ando, B. Fowler, and F. Stern, Rev. Mod. Phys. {\bf 54}, 437 (1982)

\bibitem{Luttinger60}
J. M. Luttinger and J. C. Ward, Phys. Rev. {\bf 118}, 1417 (1960);
W. Kohn and J. M. Luttinger, Phys. Rev., {\it 118} 41 (1960)

\bibitem{ggpair}
P. Gori-Giorgi, S. Moroni and G. B. Bachelet, Phys. Rev. B {\bf 70},
115102 (2004)

\bibitem{pd2d}
M. W. C. Dharma-wardana and F. Perrot., Phys. Rev. Lett. {\bf 90},
 136601 (2003)
\bibitem{prl2}
Fran\c{c}ois Perrot and M. W. C. Dharma-wardana,  Phys. Rev. Lett. {\bf 87},
 206404 (2001)
 
\bibitem{Isihara80}
A. Isihara and T. Toyoda, Phys. Rev. B {\bf 21} 3358 (1980)

\bibitem{mahan}
S. Hong and G. D. Mahan, Phys. Rev. B {\bf 52}, 7860 (1995)

\bibitem{quasi}M. W. C. Dharma-wardana, Phys. Rev. B  {\bf 72}, 125339 (2005)

\bibitem{br2d}
M. W. C. Dharma-wardana, Phys. Rev. B 82, 195303 (2010)

\bibitem{rosen}
Y. Rosenfeld and N.W. Ashcroft, Phys. Rev. A {\bf 20}, 2162 (1979)


\bibitem{Asgari09}
R. Asgari, T. Gokmen, B. Tanatar, M. Padmanabhan, and
M. Shayegan, Phys. Rev. B {\bf 79}, 235324 (2009)

\bibitem{zhang}
Y. Zhang and S. Das Sarma, cond-mat/0312565


\bibitem{kwon}
Y. Kwon, D. M. Ceperley, and R. M. Martin, Phys. Rev. B {\bf 50}, 1684 (1994)

\bibitem{Padman}
M. Padmanabhan, T. Gokmen, N. C. Bishop, and M.
Shayegan, Phys. Rev. Lett. {\bf 101}, 026402 (2008);
 T. Gokmen, M. Padmanabhan, K. Vakili, E. Tutuc, and M.
Shayegan, Phys. Rev. B {\bf 79}, 195311 (2009)

\bibitem{DrumNeeds}
N. D. Drummond and R. J. Needs, Phys. Rev. B (2009)


\bibitem{Holzmann}
 M. Holzmann, B. Bernu, V. Olevano, R.M. Martin, and D.M. Ceperley,
 Phys. Rev. B {\bf 79}, 041308(R) (2009);


\end{thebibliography}
\end{document}